On the fundamentals of Rayleigh-Taylor dynamics with variable acceleration

Aklant K. Bhowmick (1); Desmond L. Hill (1); Snezhana I. Abarzhi* (2)

Carnegie Mellon University, USA (1); University of Western Australia, AUS (2)

*corresponding author snezhana.abarzhi@gmail.com

Rayleigh-Taylor instability (RTI) has critical importance for a broad range of processes in nature and technology, from supernovae to plasma fusion. In most instances RTI is driven by variable acceleration whereas the bulk of existing studies have considered constant acceleration. This work focuses on RTI driven by acceleration with power-law time-dependence, and applies group theory to solve the classical problem. For early time dynamics, we find dependence of RTI growth-rate on acceleration parameters and initial conditions. For late time dynamics, we directly link interface dynamics to interfacial shear, find continuous family of regular asymptotic solutions, and discover invariance properties of nonlinear RTI. Our results reveal the interfacial and multi-scale character of RTI with variable acceleration. The former is exhibited in structure of flow fields with intense fluid motion near the interface and effectively no motion in the bulk; the latter follows from the invariance properties of nonlinear dynamics defined by the interplay of two macroscopic length-scales - the wavelength and the amplitude. Our theory resolves the long-standing problem of RTI nonlinear dynamics, achieves excellent agreement with observations, and elaborates diagnostic benchmarks for future experiments and simulations.

Rayleigh-Taylor instability (RTI) has critical importance in a broad range of processes in nature and technology, from celestial events to molecules [1]. Examples include supernovae, plasma fusion, fossil fuel recovery, and nano-fabrication [1-6]. RTI develops when fluids of different densities are accelerated against their density gradients and leads to intense interfacial Rayleigh-Taylor (RT) mixing of the fluids [7-9]. To advance knowledge of non-equilibrium dynamics, to better understand RT-relevant phenomena, and to positively impact developments in energy and environment, reliable theory of RTI is required [1,10]. In this work we study the long-standing problem of RTI subject to a variable acceleration and employ group theory to solve the boundary value problem for early- and late-time RT evolution [1,11,12]. We directly link (for the first time, to our knowledge) the interface dynamics to the interfacial shear, discover its invariance properties in a broad range of acceleration parameters, and reveal the interfacial and multi-scale character of RT dynamics. Extensive theory benchmarks are elaborated for experiments and simulations.

RT flows with variable acceleration commonly occur in fluids, plasmas, materials [1-12]: Blast wave driven RT mixing in a supernova enables conditions for heavy mass element synthesis. RT unstable plasma irregularities in the Earth's ionosphere lead to regional climate change. RT instabilities control ignition and burn in inertial confinement fusion. RT mixing of water and oil determines efficiency of oil recovery. RTI drives materials' transformation in fabrication of nano-particles [1-12]. These RT flows, while occurring in distinct physical circumstances, have similar features of the evolution [8-15]. RTI starts to develop when the flow fields and/or the interface are slightly perturbed near the equilibrium state [7]. the interface is transformed to a composition of small-scale shear driven vortical structures and a large-scale coherent structure of bubbles and spikes, where the bubble (spike) is a portion of the light (heavy) fluid penetrating the heavy (light) fluid [8,10-18]. Intense interfacial fluid mixing ensues with time [10,11,14-18].

RTI and RT mixing are challenging to study in experiments, simulations, theory [1,10]. RT experiments in fluids and plasmas use advanced technologies to meet tight requirements for the flow implementation, diagnostics and control [13-15]. RT simulations employ highly accurate numerical methods and massive computations to track unstable interfaces, capture small-scale processes and enable large span of scales [16-18]. In theory we have to develop new approaches to study non-equilibrium RT dynamics, identify universal properties of asymptotic solutions, and capture symmetries of RT flows [19-26]. Significant success has been recently achieved in the understanding of RTI and RT mixing with constant acceleration [1,10]. In particular, group theory approach has found a multi-scale character of nonlinear RTI and an order in RT mixing, thus explaining the observations [10-18].

Here we study RTI subject to variable acceleration. Only limited information is currently available on RT dynamics under these conditions suggesting a need in a systematic analysis [1,28]. We consider accelerations with a power-law time-dependence. On the side of fundamentals, power-law functions are important to study because they may result in new invariant and scaling properties of the dynamics [27]. As regards to applications, power-law functions can be used to adjust the acceleration's time-dependence in realistic environments and ensure practicality of our results [1-6,28].

We consider RTI in a three-dimensional spatially extended periodic flow, and apply group theory to solve the boundary value problem involving boundary conditions at the interface and at the outside boundaries [12,25,26,28]. For early-time dynamics we identify the dependence of RTI growth-rate on acceleration's parameters and initial conditions. For late-time dynamics, we directly link the interface dynamics to interfacial shear, find a continuous family of regular asymptotic solutions, and discover invariance properties of nonlinear RTI. The parameters of the critical, Atwood, Taylor and flat bubbles are identified, including their velocity, curvature, Fourier amplitudes, and interfacial shear. We reveal the essentially interfacial and multi-scale character of RT dynamics. The former is exhibited by the velocity field with intense fluid motion near the interface and effectively no motion in the bulk. The latter follows from the invariance properties of the dynamics set by the interplay of two macroscopic length-scales - the wavelength and the amplitude of the interface. We identify formal, physical, and global properties of RT flows, including universality of highly symmetric 3D dynamics and discontinuity of 3D-2D dimensional crossover. Our theory resolves the long-standing of nonlinear RT dynamics [19-26], achieves excellent agreement with observations [13-16,28,29], and elaborates diagnostic benchmarks for experiments and simulations and for better understanding of RT relevant processes in nature and technology [2-6,28,29].

RT dynamics of ideal fluids is governed by conservation of mass, momentum and energy

$$\partial\rho/\partial t + \partial\rho v_i/\partial x_i = 0,\ \partial\rho v_i/\partial t + \sum_{j=1}^{3}\partial\rho v_i v_j/\partial x_j + \partial P/\partial x_i = 0,\ \partial E/\partial t + \partial(E+P)v_i/\partial x_i = 0 \quad (1a)$$

with spatial coordinates $(x_1, x_2, x_3) = (x, y, z)$; time $t$; fields of density, velocity, pressure and energy $(\rho, \mathbf{v}, P, E)$, with $E = \rho(e + \mathbf{v}^2/2)$ and $e$ being specific internal energy [9]. Boundary conditions at the interface and at the outside boundaries are

$$[\mathbf{v}\cdot\mathbf{n}] = 0,\ [P] = 0,\ [\mathbf{v}\cdot\boldsymbol{\tau}] = any,\ [W] = any,\ \mathbf{v}|_{z\to+\infty} = 0,\ \mathbf{v}|_{z\to-\infty} = 0 \quad (1b)$$

where $[...]$ denotes the jump of functions across the interface; $\mathbf{n}(\boldsymbol{\tau})$ are the normal and tangential unit vectors of the interface with $\mathbf{n} = \nabla\theta/|\nabla\theta|, (\mathbf{n}\cdot\boldsymbol{\tau}) = 0$; $\theta = \theta(x, y, z, t)$ is a local scalar function, with $\theta = 0$ at the interface and $\theta > 0$ ($\theta < 0$) in the bulk of the heavy (light) fluid marked hereafter with sub-script $h(l)$. Specific enthalpy is $W = e + P/\rho$. Normal components of velocity and momentum are continuous at the interface, tangential velocity component and energy are discontinuous at the interface, and sources are absent at the outside boundaries [9-12].

The flow is periodic in the plane $(x,y)$ normal to the $z$ direction of acceleration $\mathbf{g}$, $|\mathbf{g}|=g$. The acceleration is directed from the heavy to light fluid, $\mathbf{g}=(0,0,-g)$ and is a power-law function of time $g=Gt^a$, where $a$ is exponent, $a \in (-\infty,+\infty)$, $G$ is pre-factor, $G>0$, and dimensions are $\dim a = 1$, $\dim G = m/s^{2+a}$. We study acceleration-driven RT dynamics with $a>-2$ [2,29]. Initial conditions include initial perturbations of the interface and flow fields with wavelength $\lambda$ [9-14]. In ideal fluids the initial conditions set length-scale $\lambda$ and time-scale $\tau \sim (G/\lambda)^{-1/(a+2)}$ of RT dynamics. In realistic fluids, small scales are usually stabilized, and scales $\lambda_\nu \sim (\nu^{(a+2)}/G)^{1/(2a+3)}$, $\tau_\nu \sim (\nu/G^2)^{1/(2a+3)}$ correspond to a fastest growing mode, where $\nu$ is the kinematic viscosity [19-21,28]. Time is $t > t_0 > 0$, $t_0 \gg \tau$. The Atwood number is $A=(\rho_h - \rho_l)/(\rho_h + \rho_l)$, $0 < A \leq 1$, $A \to 1^-(0^+)$ as $(\rho_l/\rho_h) \to 0^+(1^-)$ [7-31].

To solve the problem of RTI with variable acceleration, we employ group theory [10-12]. This approach solves the nonlinear boundary value problem and initial value problems with account for the non-local and singular character of RT dynamics, employs canonical forms of Fourier series and spatial expansions, and strictly obeys conservation laws. By using techniques of the theory of discrete groups, we first identify groups enabling structurally stable dynamics (e.g., groups of hexagon p6mm, square p4mm, rectangle p2mm in 3D; group pm11 in 2D). We next apply irreducible representations of a relevant group to expand flow fields as Fourier series, and further make spatial expansions in a vicinity of a regular point at the interface. Governing equations are reduced to a dynamical system in terms of surface variables and moments. The system's solution is sought [10-12,24-26,28].

We focus on large-scale coherent dynamics, with scales $\sim \lambda$, presuming that shear-driven interfacial vortical structures are small, with scales $\ll \lambda$ [10-12]. For convenience, all derivations are performed in the frame of reference moving with velocity $v(t)$ in the $z$-direction, where $v(t) = \partial z_0/\partial t$ and $z_0$ are the velocity and position of the bubble (spike) in laboratory reference frame. For large-scale coherent structure fluid motion is potential. For symmetry group p6mm, the velocity is $\mathbf{v}_{h(l)} = \nabla \Phi_{h(l)}$, with $\Phi_h(\mathbf{r},z,t) = \sum_{m=0}^{\infty} \Phi_m(t)\left(z + [\exp(-mkz)/(3mk)]\sum_{i=1}^{3}\cos(m\mathbf{k}_i \mathbf{r})\right) + f_h(t) + \textit{cross terms}$ and $\Phi_h \to \Phi_l$ upon $h \to l$, $\Phi_m \to \widetilde{\Phi}_m$, $z \to -z$. Here $\mathbf{r}=(x,y)$, $\mathbf{k}_i$ are the vectors of the reciprocal lattice with $\mathbf{k}_i \mathbf{a}_j = 2\pi \delta_{ij}$, $\mathbf{a}_i$ are the spatial periods in the $(x,y)$ plane, $|\mathbf{a}_i|=\lambda$, $k=|\mathbf{k}_i|=4\pi/(\lambda\sqrt{3})$, $i,j=1,2,3$, $\Phi_m(\widetilde{\Phi}_m)$ are the Fourier amplitudes of the heavy (light) fluid, $f_{h(l)}$ are time-dependent functions, and $m$ is integer. Fluid interface is $\theta = -z + z^*(x,y,t)$ with $z^* = \sum_{N=1}^{\infty} \zeta_N(t) \mathbf{r}^{2N} + cross\ terms$, where $\zeta_N$ are surface variables, $\zeta_1 = \zeta$ is the principal curvature at the bubble (spike) tip, $N$ is approximation order [10-12,25,26,28].

To the first order $N = 1$, the interface is $z^* = \zeta(x^2 + y^2)$, and dynamical system is [10-12,25]

$$\rho_h(\dot{\zeta} - 2\zeta M_1 - M_2/4) = 0,\ \rho_l(\dot{\zeta} - 2\zeta\tilde{M}_1 + \tilde{M}_2/4) = 0,\ M_1 - \tilde{M}_1 = any, \quad (2)$$

$$\rho_h(\dot{M}_1/4 + \zeta\dot{M}_0 - M_1^2/8 + \zeta g) = \rho_l(\dot{\tilde{M}}_1/4 - \zeta\dot{\tilde{M}}_0 - \tilde{M}_1^2/8 + \zeta g),\ M_0 = -\tilde{M}_0 = -v$$

where $M(\tilde{M})$ are the heavy (light) fluid moments, with $M_n(\tilde{M}_n) = \sum_{m=0}^{\infty} \Phi_n(\tilde{\Phi}_n) k^n m^n + cross\ terms$, each of which is an infinite sum of weighted Fourier amplitudes.

For early-time dynamics, $(t - t_0) \ll \tau$, $\tau = (kG)^{-1/(a+2)}$, only first order harmonics are retained in the expressions for moments $M_n(\tilde{M}_n) = k^n \Phi_1(\tilde{\Phi}_1),\ n = 0,1,2$. For a nearly flat interface the solution is $(-\zeta/k) = C_1\sqrt{t/\tau}\ I_{1/2s}(\sqrt{A}(t/\tau)^s/s) + C_2\sqrt{t/\tau}\ I_{-1/2s}(\sqrt{A}(t/\tau)^s/s),\ v = (4/k)d(-\zeta/k)/dt$ where $I_p$ is the modified Bessel function of the $p$th order, $s = (a+2)/2$, $C_{1(2)}$ are integration constants defined by initial conditions $\zeta(t_0),\ v(t_0),\ |\zeta(t_0)/k|,\ |v(t_0)\tau k| \ll 1$ [8,9,28]. Analysis of early-time dynamics, with $t \sim t_0$, $\zeta - \zeta(t_0) = -(k/4)[kv(t_0)(t-t_0)]$, $v - v(t_0) = -4A(\zeta_0/k)(\tau k)^{-1}(t_0/\tau)^a[(t-t_0)/\tau]$, suggests that positions of bubbles (spikes) with $\zeta \leq 0, v \geq 0\ (\zeta \geq 0, v \leq 0)$ are defined by initial morphology of the interface, with bubbles (spikes) formed for $\zeta(t_0)/k < 0\ (> 0)$ [28].

At late times, spikes are singular (the singularity is finite-time for $0 < A < 1$), whereas bubbles are regular [10-12]. For $t \gg \tau$ regular asymptotic solutions depend on time as $v, M, \tilde{M} \sim t^{a/2}$ and $\zeta \sim k$, and higher order harmonics are retained in the expressions for moments. Group theory is applied to solve the closure problem, find regular asymptotic solutions forming a continuous family, study the solutions' stability, and elaborate properties of nonlinear RTI [25,26,28].

For family of nonlinear regular asymptotic solutions, the bubble velocity $v \geq 0$ depends on its curvature $\zeta, \zeta < 0$, as

$$v = (\tau k)^{-1}(t/\tau)^{a/2}(-2A(\zeta/k))^{1/2}(9 - 64(\zeta/k)^2)(-48(\zeta/k) + A(9 + 64(\zeta/k)^2))^{-1/2} \quad (3a)$$

For every $A$, this function domain is $\zeta \in (\zeta_{cr}, 0),\ \zeta_{cr} = -(3/8)k$ and range is $v \in (0, v_{max})$, with $v = 0$ achieved at $\zeta = 0$ and at $\zeta = \zeta_{cr}$, and $v = v_{max}$ achieved at $\zeta = \zeta_{max}, \zeta_{max} \in (\zeta_{cr}, 0)$. The multiplicity of nonlinear solutions is due to the singular and non-local character of RT dynamics. The number of family parameters is set by the flow symmetry. For group p6mm the dynamics is highly isotropic, $z^* \sim \zeta(x^2 + y^2)$, and interface morphology is captured by principal curvature $\zeta$, Figure 1 [28].

In higher orders, $N > 1$, the solutions can be found likewise. For $N > 1$ the solutions exist and converge with increase in $N$. For $\zeta \in (-\zeta_{cr}, 0)$ the lowest-order amplitudes $|\Phi_1|, |\widetilde{\Phi}_1|$ are dominant, and values $|\Phi_m|, |\widetilde{\Phi}_m|$ decay exponentially with increase of $m$. For $\zeta \sim \zeta_{cr}$ the convergence no longer holds.

The multiplicity of nonlinear regular asymptotic solutions is due to the presence of shear at the interface, Figure 1. Defining shear $\Gamma$ as the spatial derivative of the jump of tangential velocity at the interface, $\Gamma = \Gamma_{x(y)}$, with $\Gamma_{x(y)} = \partial[v_{x(y)}]/\partial x(y)$, we find that near the bubble tip the shear is $\Gamma = -M_1 + \widetilde{M}_1$. Shear $\Gamma$ depends on the bubble curvature $\zeta$ as

$$\Gamma = \tau^{-1}(t/\tau)^{a/2}\, 12(-2A(\zeta/k))^{1/2}(-48(\zeta/k) + A(9 + 64(\zeta/k)^2))^{-1/2} \quad (3b)$$

For $\zeta \in (\zeta_{cr}, 0)$ shear $\Gamma$ is 1-1 function on $\zeta$, $\Gamma \in (\Gamma_{min}, \Gamma_{max})$ achieving $\Gamma_{min} = 0$ at $\zeta = 0$ and $\Gamma_{max}$ at $\zeta = \zeta_{cr}$. For $\zeta \in (\zeta_{cr}, 0)$ and $\Gamma \in (0, \Gamma_{max})$, the bubble velocity $v$ is a 1-1 function on the interfacial shear $\Gamma$. We present elsewhere the cumbersome function $v(\Gamma)$.

It is a common wisdom that RT evolution is accompanied by shear-driven Kelvin-Helmholtz instability (KHI). The value of shear provides the estimate for KHI growth-rate, $\omega_{KHI} \sim \Gamma$. Moreover, by directly linking the interface morphology and interfacial shear in nonlinear RTI, and by further linking both to the interface velocity, our theory finds the physics interpretation and resolve the long-standing problem of multiplicity of solutions in nonlinear RTI [23-26]. According to our results, for 3D RTI with variable acceleration, there is a continuous family of nonlinear regular asymptotic solutions, with each bubble having its own curvature, velocity and interfacial shear, Figure 1 [10-12,28].

The family has some special solutions, Figure 1. For the flat bubble with curvature $\zeta_f = 0$ the solution is $v_f = 0, \Gamma_f = 0, \Gamma_f = \Gamma_{min}$. For the critical bubble with curvature $\zeta_{cr} = -(3/8)k$, the solution is $v_{cr} = 0, \Gamma_{cr} = \tau^{-1}(t/\tau)^{a/2}\sqrt{6A/(1+A)}, \Gamma_{cr} = \Gamma_{max}$. The family has a solution which we call the 'Taylor bubble' since its curvature is the same as in Ref.[8] except for the difference in the wavevector value. For the Taylor bubble $\zeta_T = -(1/8)k$, $v_T = (\tau k)^{-1}(t/\tau)^{a/2}\sqrt{8A/(3+5A)}$, $\Gamma_T = \tau^{-1}(t/\tau)^{a/2}(3\sqrt{2A/(3+5A)})$; also $(\Phi_2)_T = 0$ at $N = 1$. The fastest solution in the family $(\zeta_{max}, v_{max})$ obeys conditions $\partial v/\partial \zeta = 0, \partial^2 v/\partial \zeta^2 < 0$. We call this solution the 'Atwood bubble' to emphasize its complex dependence on the Atwood number: $\zeta_A = \zeta_{max}$, $v_A = v_{max}$. For $A \to 1^-$ the values are $\zeta_A \approx -(k/8)(1-(1-A)/8)$, $v_{max} \approx (\tau k)^{-1}(t/\tau)^{a/2}(1-3(1-A)/16)$. For $A \to 0^+$ the values are $\zeta_A \approx -(3k/16)A^{1/3}$, $v_A \approx (\tau k)^{-1}(t/\tau)^{a/2}(3/2)^{3/2}\sqrt{A}$. For $v \sim v_A$ velocity has steep dependence on shear $\Gamma$, Figure 1. This suggests the need in highly accurate methods of numerical modeling and experimental diagnostics of RT dynamics.

Analysis of asymptotic stability of solutions, with small departures $\sim \Delta(\beta,(t/\tau))$ and (un)stable solutions for $(\text{Re}[\beta] > 0) \text{Re}[\beta] < 0$, suggests that flattened bubbles are unstable, curved bubbles are stable, and the fastest stable solution $\zeta \sim \zeta_A, v \sim v_A, \Gamma \sim \Gamma_A$ is the physically significant solution, Figure 1. The solution has remarkable invariant property $v_A^2(\tau k)^{-2}(t/\tau)^{-a}(8|\zeta_A|/k)^3 = 1$ [10-12,26]. This invariance implies that nonlinear coherent RT dynamics is essentially multi-scale, with the two macroscopic scales contributing - the wavelength and amplitude [10-12,26,28]. RT coherent structure is a standing wave, whose nonlinear dynamics is defined by the growing amplitude and with the wavelength.

For fluids with very different densities $0 << A \leq 1$ parameters of the Atwood and Taylor bubbles are close to one another. At $a = 0$ and $0 << A \leq 1$ velocities $v_{A(T)}$ are also close to that of the so-called Layzer-type bubble $v_L = (\tau k)^{-1}\sqrt{2A/(1+A)}$, with which experiments and simulations usually compare well [10-12,24-26,30,31]. Thus, our results excellently agree with existing observations [13-18,28]. Our theory is focused on large-scale dynamics and presumes that interfacial vortical structures are small-scale. This assumption is applicable for fluids very different densities and with finite density ratios. For fluids with very similar densities $0 \approx A << 1$ other approaches can be employed [10-12,25,26,28].

By accurately accounting for the harmonics interplay and systematically connecting the interface velocity to interfacial shear in a broad range of acceleration parameters, we find that RT dynamics is essentially interfacial: It has intense fluid motion in a vicinity of the interface, effectively no motion away from the interface and shear-driven vortical structures at the interface, Figure 2. This velocity pattern is observed in experiments and simulations, in excellent agreement with our results [13-18,28].

We further apply group theory to find solutions for other symmetries in 3D flows and for 2D flows. 3D flows tend to conserve isotropy in the plane, and 3D highly symmetric dynamics is universal, with p6mm solutions transformed to p4mm solution upon the substitution $k \to 2\pi/\lambda$. This suggests that for given wavelength and acceleration parameters, bubbles move faster and are more curved for 'square' group p4mm when compared to group of hexagon p6mm. For 3D low symmetric dynamics with group p2mm, there is a two-parameter family of regular asymptotic solutions; among the family solutions only nearly isotropic bubbles are stable; the anisotropic bubbles are unstable (and can break or merge to keep isotropy in the plane normal to acceleration); the dimensional 3D-2D crossover is discontinuous [10-12,25,26,28]. For 2D flow with group pm11, there is a one-parameter family of regular asymptotic solutions. In 3D and in 2D velocities (curvatures) corresponding to the Atwood, Taylor, and critical bubbles relate as $\sqrt{3}$ $(3/4)$, and in 2D $v_A^2 (3\tau k)^{-2} (t/\tau)^{-a} (2|\zeta_A|/k)^3 = 1$.

For a broad range of acceleration parameters, the interface velocity is given by standard functions at early time $\sim \sqrt{t/\tau} \, I_{\pm 1/(a+2)} \left( 2\sqrt{A} \, (t/\tau)^{(a+2)/2} /(a+2) \right)$, and at late time $\sim (t/\tau)^{a/2}$, $\tau = (kG)^{-1/(a+2)}$. This property enables a comparative study of RT dynamics for various acceleration exponents and strengths, $a > -2, G > 0$. For $(-2 < a < 0)\, a > 0$, one expects a (sub)super-exponential dynamics at early time and (de)accelerated dynamics at late time, whereas nonlinear bubble velocity and shear, when rescaled as $v(\tau k)(t/\tau)^{-a/2}$ and $\Gamma\tau(t/\tau)^{-a/2}$ depend only on the interface morphology and flow symmetry. Hence, by analyzing properties of RT bubbles for fast dynamics and large exponents $a > 0$, one can obtain properties of those for slow dynamics and small exponents $-2 < a < 0$. Such derivations are especially convenient for studies of RTI in high energy density plasmas in astrophysics or fusion, where RTI is driven by an explosion or an implosion with blast waves acceleration exponents $-2 < a < -1$ [27-29].

In addition to interface velocity commonly diagnosed in experiments and simulations [13-18], our analysis elaborates the diagnostic quantities which have not been discussed before. These are the fields of velocity and pressure, the interface morphology and the bubble curvature, the interfacial shear and its link to the bubble velocity and curvature, the spectral properties of velocity and pressure, as well as the interface growth and growth-rate. By diagnosing dependence of these quantities on the density ratio, the flow symmetry, and the acceleration's exponent and strength, by identifying their universal properties, and by accurately measuring departures of data in real fluids from theoretical solutions in ideal fluids, one can further advance knowledge of RT dynamics in realistic environments, achieve better understanding of RT relevant processes, from supernovae to fusion, and improve methods of numerical modeling and experimental diagnostics of interfacial dynamics in fluids, plasmas, materials.

To conclude, we have solved the long-standing problem of RTI with variable acceleration by applying group theory. We have directly linked the interface velocity to interfacial shear, revealed the interfacial and multi-scale character of late-time RT dynamics, achieved excellent agreement with available observations, and elaborated new theory benchmarks for future experiments and simulations.

**Acknowledgements**
SIA thanks for support the University of Western Australia, AUS; the National Science Foundation, USA.

**Figure captions**

Figure 1: One parameter family of regular asymptotic solution for 3D flow with group p6mm at some Atwood numbers. Bubble velocity versus (a) bubble curvature; (b) interfacial shear; (c) interfacial shear versus bubble curvature; (d) solutions stability.

Figure 2: Qualitative velocity field in laboratory reference frame near the bubble tip at some Atwood number for a solution in RT family in (a) the volume; (b) the plane (dashed curve marks the interface).

Figure 1

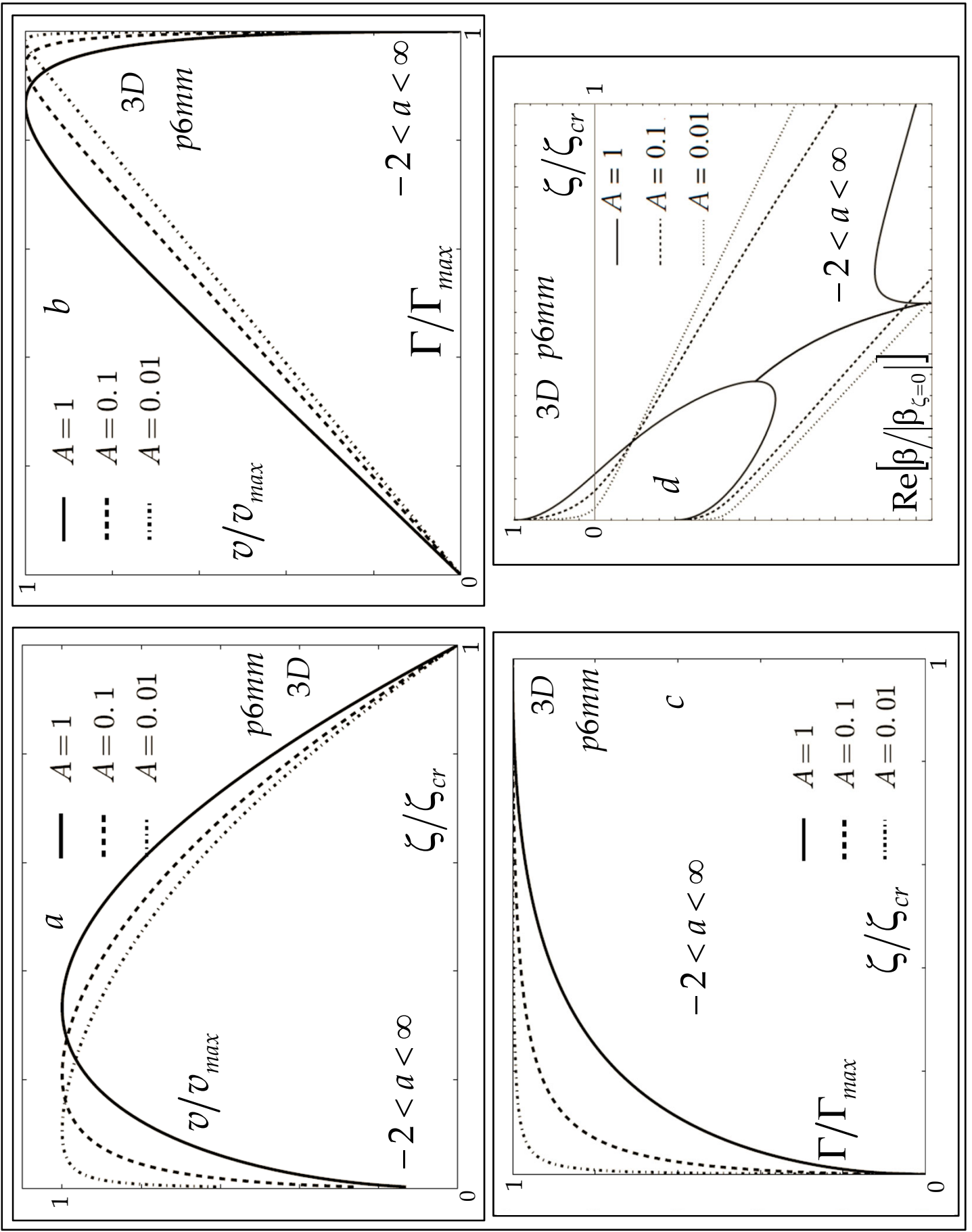

Figure 2

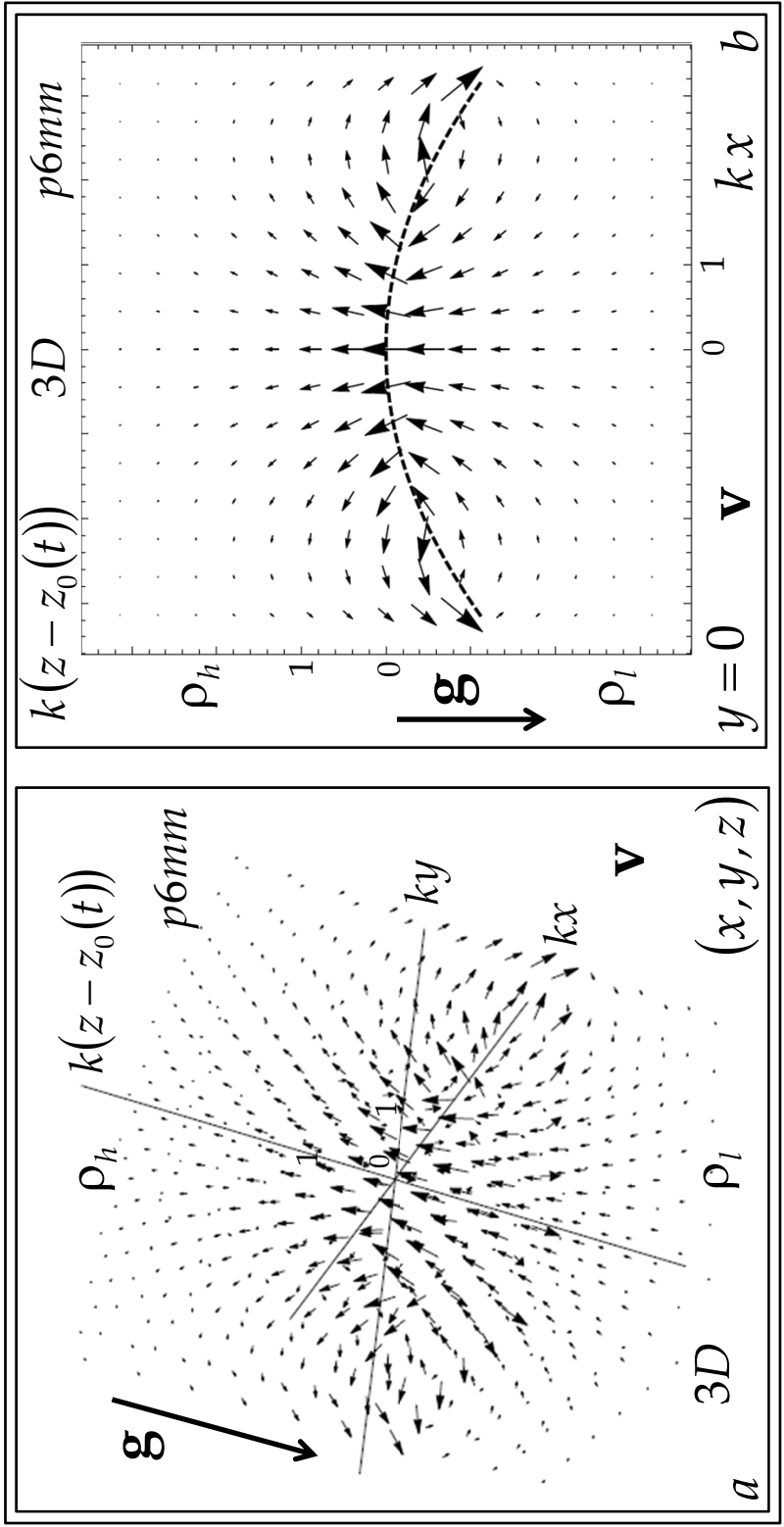